\documentclass[twocolumn, final]{svjour3} 
\RequirePackage{fix-cm}

\usepackage{graphicx}
\usepackage{dcolumn}
\usepackage{bm}
\usepackage[cmex10]{amsmath}
\smartqed 

\usepackage[utf8]{inputenc}
\usepackage{color}
\definecolor{dgreen}{rgb}{0,0.6,0}

\usepackage{stfloats}
\usepackage{url}

\newcommand{\tn}[1]{\textnormal{#1}}
\newcommand{\ee}[0]{\mathrm{e}}
\newcommand{\dd}[0]{\mathrm{d}}
\newcommand{\sub}[1]{\ensuremath{_\mathrm{#1}}}

\newcommand{\change}[1]{#1}

\begin{document}

\title{Empirical transport model of strained CNT transistors used for sensor applications}

\author{Christian~Wagner \and Jörg~Schuster \and Thomas~Gessner}

\institute{C. Wagner \and T. Gessner \at 
			Technische Universität Chemnitz,
			Center for Microtechnologies,
			Reichenhainer Straße 70,
			09126 Chemnitz\\
			\email{christian.wagner@zfm.tu-chemnitz.de}
\and J. Schuster \and T. Gessner \at
			Fraunhofer Institute for Electronic Nano Systems (ENAS),
			Techologiecampus 3,
			09126 Chemnitz}

\date{Published Online: 20 April 2016}

\maketitle

\begin{abstract}
We present an empirical model for the near-ballistic transport in carbon nanotube (CNT) transistors used as strain sensors. This model describes the intrinsic effect of strain on the transport in CNTs by taking into account phonon scattering and thermally activated charge carriers. As this model relies on a semiempirical description of the electronic bands, different levels of electronic structure calculations can be used as input. The results show that the electronic structure of strained single-walled CNTs with a radius larger than 0.7\,nm can be described by a fully analytical model in the sensing regime. For CNTs with smaller diameter, parameterized data from electronic structure calculations can be used for the model. Depending on the type of CNTs, the conductance can vary by several orders of magnitude when strain is applied, which is consistent with the current literature. Further, we demonstrate the tuning of the sensor by an external gate which allows shifting the signal amplitude and the strain sensitivity. These parameters have to be balanced to get good sensing properties. Due to its basically analytical nature, the transport model can be formulated as a compact model for circuit simulations.

\end{abstract}
             
\section{Introduction}

For the construction of novel, nanoscopic strain sensors, carbon nanotubes (CNTs) are ideal candidates. Conventional silicon-based sensors suffer from a low signal-to-noise-ratio for mechanical high-frequency applications in the MHz regime --- due to their stiff design in combination with a capacitive readout. CNTs have the abilitiy to change their bandgap upon strain \cite{Yang_2000,Kleiner_2001,Leeuw_2008,Wagner_2012b}. This opens the possibility of strain-engineering in CNTs. For mechanical sensor applications, this results in a drastic change of resistance \cite{Minot_2003,Stampfer_2007}, such that they show an enormous sensitivity. In addition, low device capacitances in the order of $10^{-15}$\,F enable high frequency operation of CNT based sensors (for a device containing approximately 100 CNTs with 1\,$\mu$m length). CNTs are also mechanically stable (Young's modulus: 1\,TPa, fracture strength: $\approx$ 100\,GPa) and can be strained up to 10\% without mechanical failure \cite{Wu_2008,Jeong_2010,Parvaneh_2010,Wagner_2012}, which makes them very reliable for this application.



The design of CNT-based sensors in a device structure is not straight forward so that there is need for drafting and simulation tools. For circuit and sensor system simulation such tools need to be of low computational cost. Thus, compact models are often chosen for this task. Compact models are models that contain parameterized relations describing the behavior of a subsystem under external influences. These models can be based on experimental observations as well as physics-based simulations. 

\begin{figure}[t] 
\includegraphics[width=84mm]{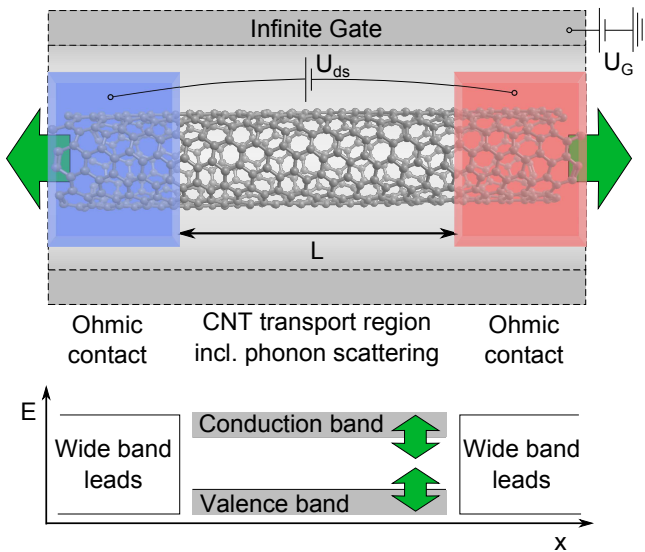}
\caption{A sketch of the simplified CNT device: A quasi infinite, strained CNT with ideal, ohmic contacts. An according band diagram is shown below.} \label{fig_device}
\end{figure}

There are already different compact models available for non-strained CNT transistors \cite{Maneux_2013,Schroter_2013,Luo_2013}, which are still under discussion \cite{Claus_2012}. We focus on intrinsic piezoresistive transport properties of CNTs with disregard of contact effects. This approach is called Pseudo-bulk-approximation in \cite{Schroter_2013} and it is illustrated in in figure \ref{fig_device}. Our model bases on the following assumptions:

\begin{itemize}
 \item A parameterized description of conduction and valence bands of the CNT 
 \item Near-ballistic transport based on the model of Zhou~\cite{Zhou_2005} 
 \item Finite length of the CNT with ideal, ohmic contacts (no contact scattering)
 \item Homogenous gate throughout the device (which is mostly true as common device setups have an extended back-gate or a wrap-around-gate \cite{Steiner_2012,Shulaker_2013,Ning_2014})
 \item Thermal equilibrium distribution of charge carriers in the CNT
 \item Finite bias with1in the linear response regime
\end{itemize}

The above-mentioned compact models include scattering at contacts, but they are missing axial strain as input. Contact effects in a strained CNT device, however, are not included in a straight forward way as the band alignment between the CNT and the contacts shifts upon strain --- and influences the device sensitivity. The amount of this shift is a priori unknown and therefore, our study is restricted to intrinsic effects.

Still, our model gives insights into device performances as well as the desired operation regime. Thus, it goes far beyond the first, fully ballistic transport approximation of Minot \cite{Minot_2003}, which is often used in literature to approximate the performance of CNT strain sensors \cite{Stampfer_2007,Cao_2003,Grow_2005,Cullinan_2010}. 

Bipolar CNT transistors, contacted with Pd, can show symmetric transfer characteristics, which means that they show a small or no Schottky barrier \cite{Ning_2014,Chen_2005,Svensson_2009} --- in those situations, our model is accurate, even if the contacts are not considered. 

\section{Computational model for near-ballistic transport in CNT transistors at finite temperatures}

The goal of our computational model is to be numerically cheap, but still accurate --- therefore, we need a simplified, but exact description of the CNT physics. Thus, the electronic structure and the transport problem of the strained CNTs need to be treated in a parameterized way.

We rely on the model for transport in CNT transistors sketched by Zhou \cite{Zhou_2005}. For this purpose, an analytical description of the CNT electronic band structure is preferred. The electronic structure of CNTs is quite well understood \cite{Anantram_2006,Charlier_2008,Bellucci_2011} and one can model it at different levels of sophistication: Electronic structure methods like density functional theory (DFT), tight binding (TB) methods or the even simpler Mintmire approximation \cite{Mintmire_1998}. We compared the different levels of sophistication systematically in an earlier work and found that they can be brought into agreement for strained CNTs with radii $r > 3.5\,$\AA{} \cite{Wagner_2012b}. Regarding CNTs with smaller radii, electron structure calculations using DFT will be required to capture the effects of strong bending on the electronic structure which will be no longer similar to graphene \footnote{The parameters for density functional calculations are identical to those published in \cite{Wagner_2012b}}.

The description of a CNT band is based on a cone section and therefore needs three independent parameters, namely the effective mass ($m^\ast$), the Fermi velocity of the underlying graphene-TB-model ($v_0\approx 8.6\cdot 10^5\, \tn{ms}^{-1}$ for the first CNT sub-band) and a bandgap correction for valence and conduction band ($\Delta E\sub{B}$):

\begin{equation}
E(k) = \pm\sqrt{(\hbar \change{(k-k')}{\color{blue} v_0})^2 + ({\color{blue} m^\ast v_0}^2)^2} + {\color{blue} \Delta E\sub{B} }
\label{equ_bands}
\end{equation}

The bandgap correction $\Delta E\sub{B}$ is obtained by fitting the model against bands from electronic structure methods. It turns out to be close to zero for the first CNT sub-band, but for other sub-bands, it is non-vanishing. \change{Further, we introduced a shift $k-k'$ instead of $k$ in order to fit band minima, which are not at the Gamma point} \footnote{\change{This fact seems unlikely for a material, but already the Tight binding zone folding approach predicts a band minimum for chiral CNTs, which is not at the Gamma point.}}. \change{However, this shift has no further influence on the results and will be disregarded.}

\begin{figure}[t] 
\includegraphics[width=84mm]{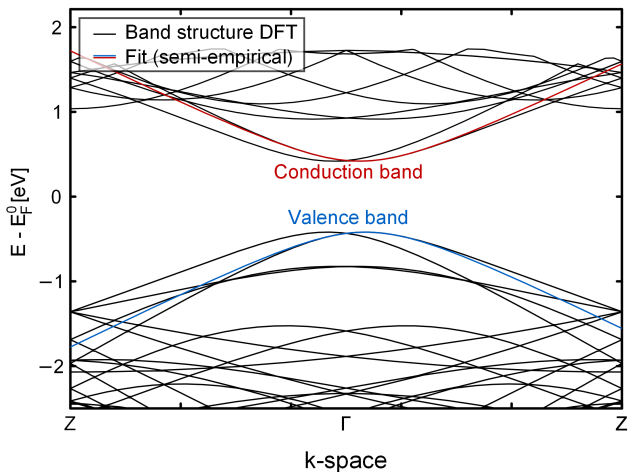}
\caption{The band structure of the (8,4)-CNT at zero strain: The result obtained by DFT is shown and the effective band structure has been fit against the data. Conduction band:$m^\ast\sub{c}=0.097\,\mathrm{m_{el}}$ $v_0=863'000\,\mathrm{ms^{-1}}$ $\Delta E \sub{B,c}=0.009\,$ Valence band: $m^\ast\sub{v}=0.097\,\mathrm{m_{el}}$ $v_0=884'000\,\mathrm{ms^{-1}}$  $\Delta E\sub{B,v}=0.012\,$eV.} \label{fig_bands}
\end{figure}

An example of this fit to DFT data of an (8,4)-CNT is shown in figure \ref{fig_bands}. $E\sub{F}^0$ denotes the Fermi level, when no Gate voltage is applied. The chiral (8,4)-CNT is taken as an example throughout the whole publication as it is semiconducting with a finite bandgap and has a radius, which is comparable to experimentally available CNTs, e.g. \cite{Leeuw_2008}. With respect to the computational effort its unit cell is small enough to treat the system by DFT. The chirality of the (8,4)-CNT is also beneficial as it is not a high symmetry situation which might hide errors in the analytical description.

In the following, the electronic transport model is described. We use the Landauer-Büttiker-formalism \cite{Bagwell_1989} to calculate the current through the CNT at a given Fermi level $E\sub{F}$:

\begin{align} \label{equ_land}
I(E\sub{F}) = G\sub{0}\!\int\limits_{-\infty}^{\infty} \!\!\mathcal{T}(E-E\sub{F})\cdot (f(E\sub{-} -f(E\sub{+}))\,\tn{d}E \\
\tn{with } f(E) = \frac{1}{1+\ee^{E/k\sub{B} T}} \tn{ and } E\sub{\pm} = E\pm\frac{U\sub{ds}}{2}. \nonumber
\end{align}

$f(E)$ describes the Fermi function at a given Temperature $T$, $G\sub{0}=2\frac{e^2}{h}$ the conductance quantum and $U\sub{ds}$ the drain-source-voltage. Here, $\mathcal{T}(E)$ can be any transmission spectrum. The transmission spectrum of an ideal CNT without scattering is zero within the bandgap and two within a band. In the limit of $T=0$\,K and at zero bias, this formula simplifies to $I(E)=U\sub{ds}\mathcal{T}(E)G\sub{0}$, which is the basis of the model of Zhou \cite{Zhou_2005}. The quantity $G(E)$ can be referred as the conductance spectrum 

\begin{equation}
G(E) = 1/{U\sub{ds}} \cdot I(E).
\end{equation}

Formally, equation \ref{equ_land} is equivalent to a convolution of the conductance spectrum at zero Kelvin $G(E)$ and the product of the Fermi functions at the contacts. The drain-source-voltage $U\sub{ds}$ is applied symmetrically to keep the Off-state of the transistor at a Gate voltage of $U\sub{G}=0\,$V -- which is not a restriction of the model. In real devices, the prediction of the off-state is not straight-forward and usually considered as a fit parameter \cite{Claus_2012}.

\begin{figure*}[t]
 \includegraphics[width=174mm]{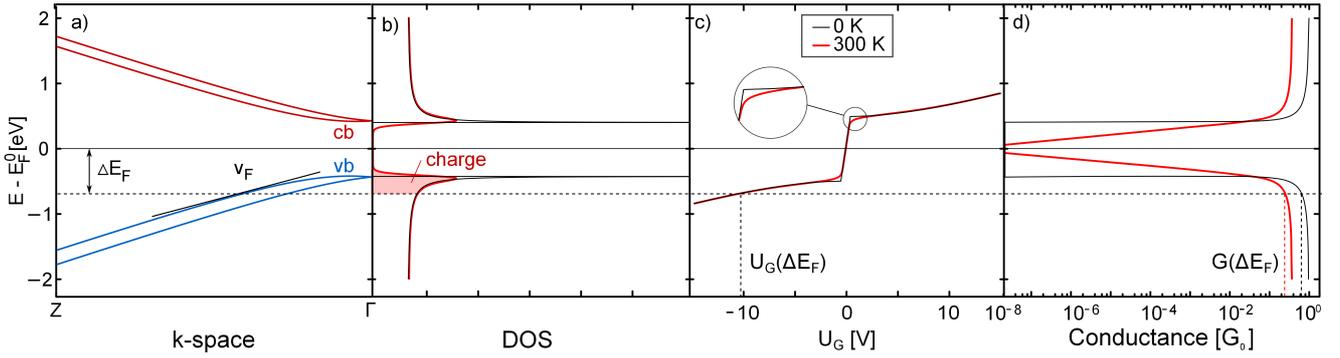}
 \caption{Overview of the transport model: a) Simplified band structure of the CNT according to equation \ref{equ_bands} (drawn in in figure \ref{fig_bands}) with the Fermi velocity of the carriers corresponding to a given Fermi energy $E\sub{F}$. b) Density of states (DOS) of the CNT at 0\,K and broadened with a 300\,K Fermi distribution function. The charge stored in the CNT is filled in red. c) The dependence of $E\sub{F}$ on the applied gate voltage. d) Conductance spectrum of the model (0\,K \cite{Zhou_2005} without scattering) and broadened by a Fermi function at 300\,K.} \label{fig_compact}
\end{figure*}

Phonon scattering can be taken into account by a modified CNT transmission spectrum $\mathcal{T}\sub{scat}(E)$: 

\begin{align}\label{equ_cond}
\mathcal{T}\sub{scat}(E) &= \mathcal{T}(E) \overbrace{\frac{\ell\sub{F}(E)}{L+\ell\sub{F}(E)}}^{|t|^2(E)},
\end{align}

$\ell\sub{F}$ is the mean free path (MFP) and $L$ is the CNT length. We use a value of $L=1000$\,nm, which is typical for CNT devices. The Matthiessen rule is applied, such that the total scattering rate reads as $|t|^2(E) = \frac{\ell\sub{F}(E)}{L+\ell\sub{F}(E)}$ -- considering the case $\ell\sub{F} > L$, the ballistic limit naturally comes out. This is a minor modification of the model of Zhou, where $|t|^2(E) = \frac{\ell\sub{F}(E)}{L}$ -- this formulation holds for $\ell\sub{F} \ll L$, which is usually the case.

The MFP is calculated by $\ell\sub{F}(E)=v\sub{F}(E)\tau\sub{F}(E)$ in a Drude approach, where $\tau\sub{F}(E)$ is the carrier scattering time and $v\sub{F}(E)$ is the Fermi velocity of a carrier at a certain energy $E$. 
 
The Fermi velocity is defined by 

\begin{equation}
v\sub{F}(E)= \left. -\frac{1}{\hbar}\frac{\partial E(k)}{\partial k} \right\vert_{k=k\sub{F}} = \frac{\hbar v_0^2 k}{E(k)}
\end{equation}

 and shown in figure \ref{fig_compact}a. $k\sub{F}$ is the Fermi wave vector at the Fermi level $E\sub{F}$. The calculation of $E\sub{F}$ is discussed below. 

The scattering time $\tau\sub{F}$ is obtained by the \change{scaling} relation\footnote{\change{The scaling relation can be obtained by Fermi's golden rule (transmission rate scales with the final states' DOS) and the scaling of the density of states, which scales with $\tau\sim 1/v$.}} $\frac{v\sub{F}}{\tau\sub{F}} = \frac{v_0}{\tau_0}$ with the relation $\tau_0 = \frac{2\hbar}{3 m^\ast \alpha T v_0}$ \cite{Ando_2002,Zhou_2005}. The parameter $\alpha$ can be taken as 9.2\,m/(Ks) \cite{Zhou_2005} and $T$ denotes the temperature. Quantities with index 0 are \change{in the limit of large k-values. Now}, we can calculate the conductance spectrum $G(E\sub{F})$ (figure \ref{fig_compact}d).

Figure \ref{fig_compact} illustrates the interplay of the different parts of the model when a finite gate voltage is applied to the CNT. It is electrostatically doped in a way that CNT states are populated. This leads to a shift of the Fermi level $E\sub{F}$ -- according to the gate capacitance per unit length $C_G'$ (figure \ref{fig_compact}b+c). The capacitance depends on the device geometry and is taken as \mbox{$1.5\cdot 10^{-11}\,$F/m} like in \cite{Zhou_2005}. Its exact value has an impact on the magnitude of this shift. Therefore, the exact value is of relevance, only, when results are compared to real devices. The charge stored in a CNT is determinied by its density of states $g(E)$ (DOS): 

\begin{equation}
g(E) = \frac{E}{\hbar v_0 \sqrt{E^2 - (m^\ast v_0^2)^2}}
\end{equation}

As we account for thermally activated carriers, the density of states needs to be convolved by a Fermi difference function $\tilde{f}$ (to be normed), which is the effective (or smeared) DOS $g_{th}$:

\begin{align}
g\sub{th}(E) &= \int\limits_{-\infty}^{\infty} g(E-E')\,\tilde{f}(E') \,\dd E' \nonumber \\ 
\tn{with }  \tn{ and } \tilde{f}(E) & =\frac{\dd f(E)}{\dd E}
\end{align}

This is valid for equilibrium transport at zero bias. One can account for finite drain-source voltage $U\sub{ds}$ within the linear response regime by modifying $\tilde{f}$. Assuming that the bias is applied symmetrically, $\tilde{f}$ results as 

\begin{align}
\tilde{f}(E, U\sub{ds}) = & \frac{1}{N}\left[ f(E-\frac{eU\sub{ds}}{2}) -f(E+\frac{eU\sub{ds}}{2}) \right] \nonumber \\
\text{with } N = & \int\limits_{-\infty}^{\infty} \!\!f(E-\frac{eU\sub{ds}}{2}) -f(E+\frac{eU\sub{ds}}{2})\,\dd E. \label{equ_lin_response}
\end{align}

The symmetric application of the drain-source-voltage has to be consistent with equation \ref{equ_land}.

Now, we can determine the position of the Fermi level at a given gate voltage $U\sub{G}(E\sub{F})$ (figure \ref{fig_compact}c). The carriers in the Gate are balanced by carriers in the CNT as a counter charge populating its states. The integrated density of states equals the number of carriers in the CNT, which then leads to an expression for the gate voltage $U\sub{G}(E\sub{F})$:

\begin{equation}
U\sub{G}(E\sub{F}) = \frac{e}{C'\sub{G}} \overbrace{\int\limits_{E\sub{F}^0}^{E\sub{F}} g\sub{th}(E)\,\dd E}^{\rho\sub{CNT}/e}
\end{equation}

$\rho\sub{CNT}$ is the charge density stored in the CNT (in this case: charge per length, as a CNT is a quasi-one-dimensional conductor) as illustrated in fig. \ref{fig_compact}b. This integral can be approximated as shown in \cite{Maneux_2013,Leonard_2006}.

\begin{figure}[t] 
 \includegraphics[width=84mm]{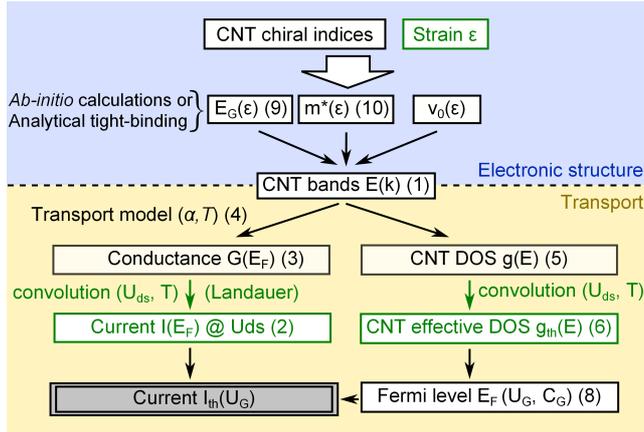}
 \caption{Workflow of the present compact model: The electronic structure problem of strained CNTs is solved, first. Upon this basis, the electron transport is calculated. Numbers in brackets give the according equation numbers in the publication. Green declarations mean that the implementation is new with respect to cited literature.} \label{fig_compact_2}
\end{figure}

Figure \ref{fig_compact}d shows the dependence of the conductance spectrum at zero Kelvin and at room temperature. The first case is a reproduced result of Zhou \cite{Zhou_2005}: When the Fermi level lies in the middle of the bandgap, the conductance drops to zero -- within a band, it is finite. To account for thermally activated carriers, the Landauer formula (equation \ref{equ_land}) needs to be applied: 

The consequence is that the conductance is still larger than zero when the Fermi level lies in the bandgap, but it drops exponentially in energy. The resulting convolution integral is solved numerically as it cannot be solved in a closed form like the one before. But analytical expressions should be possible by approximating the integral kernel similar to \cite{Maneux_2013,Leonard_2006}.


The work flow of this model is summarized in figure \ref{fig_compact_2}. For the study of transport in strained CNTs, we need their electronic structure described in the following. 

\section{Electronic structure of strained CNTs}

The basic sensor mechanism is the change of the bandgap of CNTs under strain, but also the strain dependence of the effective mass and the Fermi velocity play a role.

\begin{figure}[t] 
\includegraphics[width=84mm]{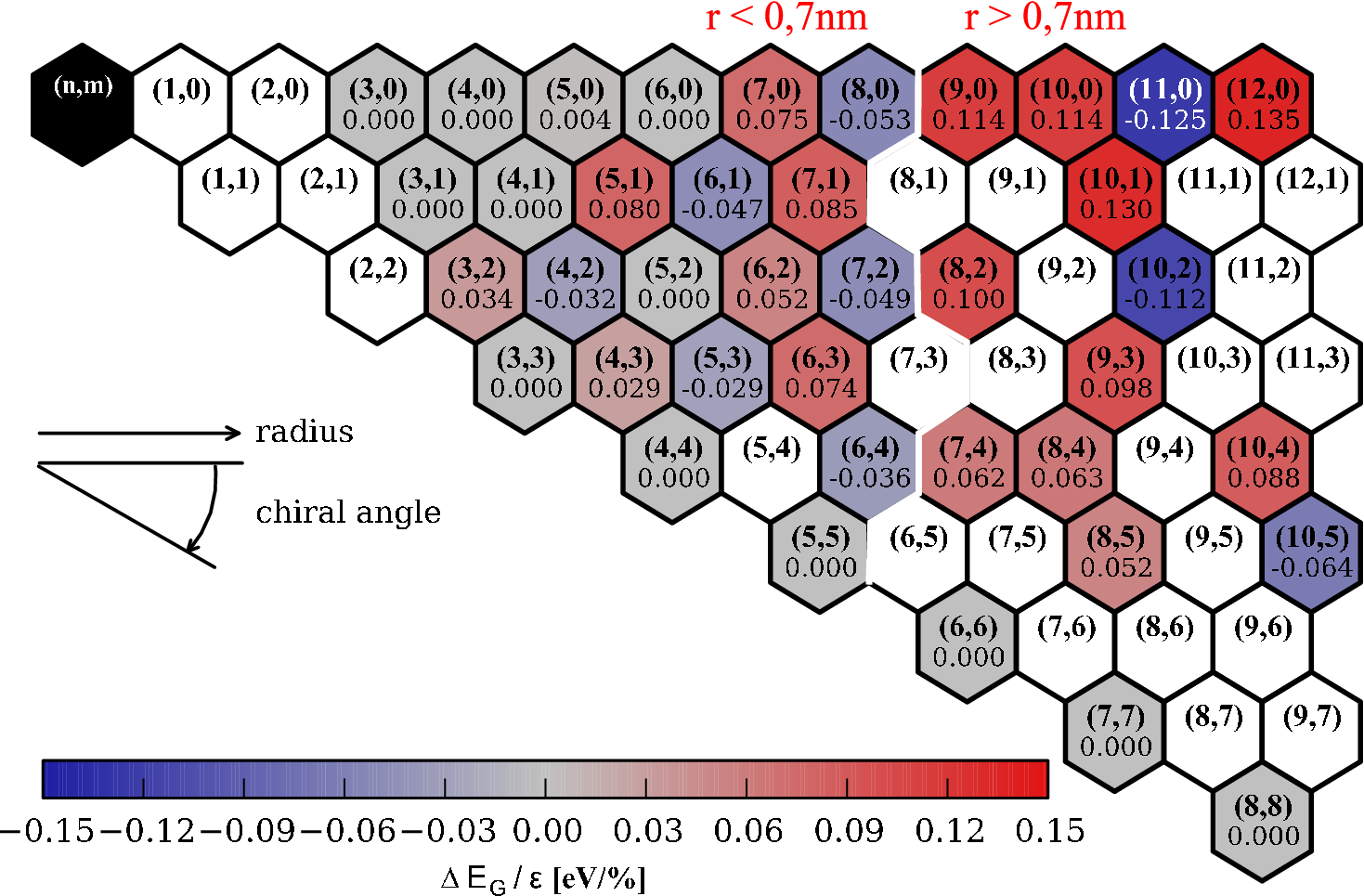}\newline
\includegraphics[width=84mm]{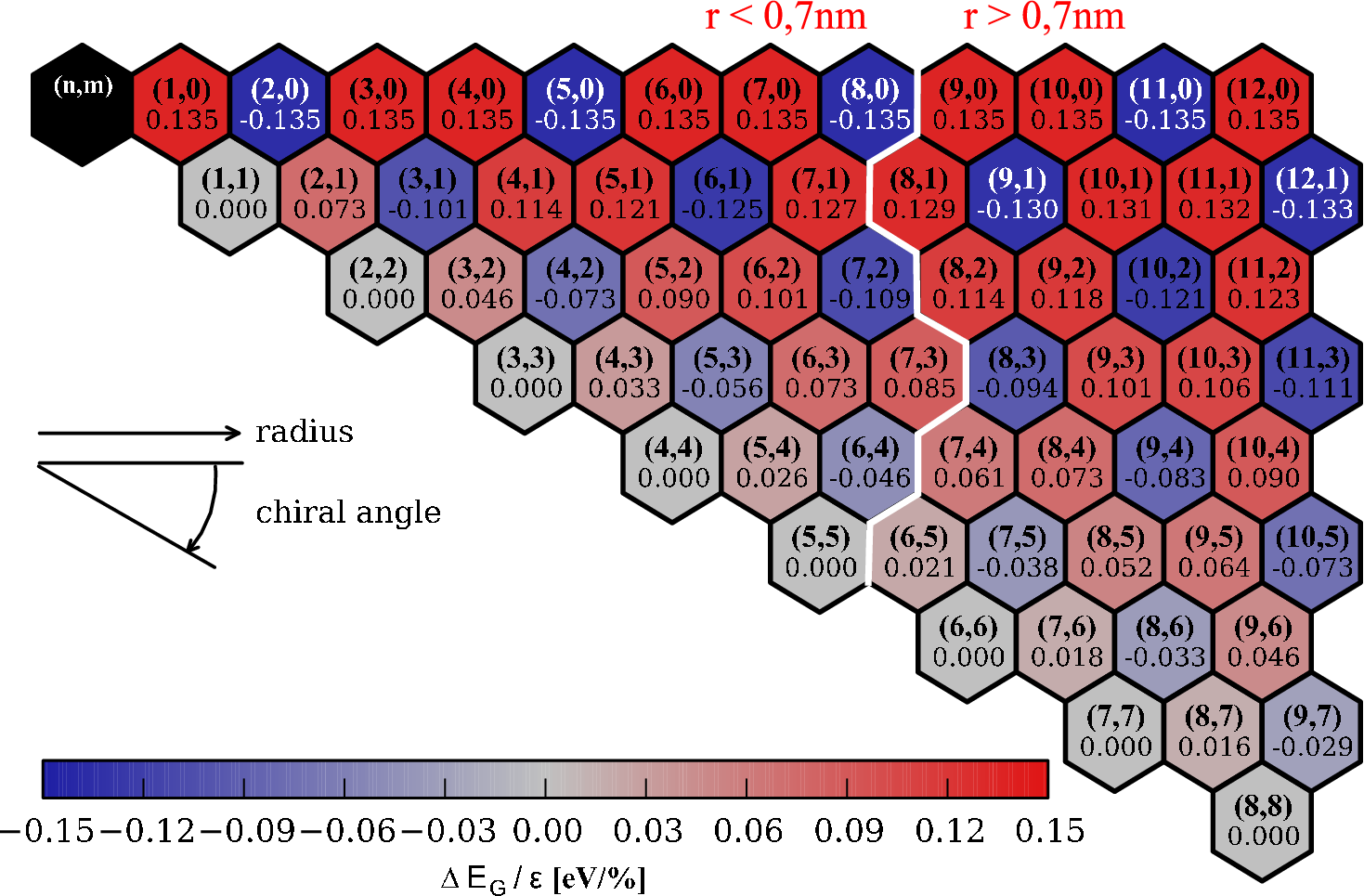}
\caption{The change of the bandgap of different CNTs with respect to positive strain: DFT data (upper image, white boxes indicate CNTs without underlying DFT data) and data of the analytical model (lower image). Both data agree sufficiently for CNTs with a diameter $d$ larger than 0.7\,nm.} \label{fig_CNT_periodic_table}
\end{figure}

The effect of strain on the bandgap can be described either analytically \cite{Yang_2000,Kleiner_2001,Wagner_2012b} or by electronic structure calculations \cite{Wagner_2012b,Valavala_2008,Sreekala_2008}. The analytical expression for the bandgap change $\Delta E_\text{G}/\varepsilon$ is the following:

\begin{equation}
\Delta E\sub{G}/\varepsilon = \tn{sgn}(2p+1)\sqrt{2}\cdot 3t_0\,\left[\,(1+\nu)\,\cos 3\theta \,\right].
\end{equation}

Here, $p = [n-m]_3$ with $p \in \{-1,0,1\}$ and $n,m$ are the chiral indices of a CNT. $t_0$ is the TB-hopping-parameter ($\approx 2.66\,$eV), $\nu$ declares the Poisson's ratio, $\varepsilon$ denotes the strain, and $\theta$ represents the chiral angle. This is exemplarily shown in figure \ref{fig_effmass} (inset) for the (8,4)-CNT in comparison to DFT data. Further theory and results concerning bandgaps are published in \cite{Wagner_2012b}.

The systematic comparison of the models is shown in figure \ref{fig_CNT_periodic_table}, where the change of the bandgap with respect to positive strain is depicted. Blue color means bandgap closing upon strain whereas red color means opening of the bandgap. White hexagons indicate CNTs where full DFT calculations are too expensive. Numerical values are shown for the quantitative comparison of DFT results (upper figure) to analytical tight-binding results (lower figure). It can be seen once more that both approaches agree nicely as long as the CNT radius is larger than 0.7\,nm and deviations occur due to curvature effects. From a practical point of view SWCNTs in experiments possess a diameter in the range of 1 -- 2.5\,nm \cite{Leeuw_2008,Minot_2003,Wu_2008} so that the analytical model holds for most relevant cases.

\begin{figure}[t] 
\includegraphics[width=84mm]{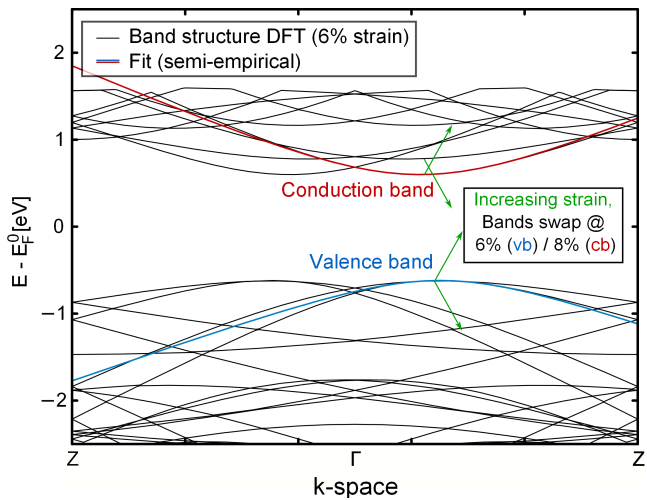}
\caption{The band structure of the (8,4)-CNT at 6\% strain: DFT results and the empiric fits are shown. The valence bands are about to swap, green arrows indicate the movement of the bands upon strain. (conduction band: $m^\ast=0.155\,\mathrm{m_{el}}$, $v_0=821'000\,\mathrm{ms^{-1}}$, $\Delta E\sub{B}=0.004\,$eV, valence band: $m^\ast=0.162\,\mathrm{m_{el}}$, $v_0=674'000\,\mathrm{ms^{-1}}$, $\Delta E\sub{B}=0.2\,$eV).} \label{fig_bands_swap}
\end{figure}

As the CNT bands are bent upon strain, the effective mass of the charge carriers is strain dependent --- in the conduction band as well as in the valence band. This has already been seen by Sreekala for the (13,0)-CNT \cite{Sreekala_2008} in a very similar way. A comparison of our results can be found in the supporting information. If one derives the carriers' effective masses from the Dirac-cone approximation of the graphene band structure, they directly depend on the bandgap. As linear axial strain does (approximately) not deform the according Dirac-cone and just shifts the cutting k-lines in the CNT k-space \cite{Yang_2000}, this formula also holds for CNTs under strain:

\begin{equation}
m^\ast(\varepsilon) = \frac{2\hbar^2}{9a_0^2t_0^2} E\sub{G}(\varepsilon).
\end{equation}

The lattice constant of graphene is given by $a_0=2.406\,$\AA{}. 

As well as the other approximations, with the given parameter set, this holds for the first sub-band, only. At a critical strain $\varepsilon^\ast$, the CNT sub-bands swap their positions with respect to the Fermi level. For the (8,4)-CNT this occurs at about 6\% strain (see figure \ref{fig_bands_swap}). Above this critical strain value, we expect deviations of the models. This can be exemplarily seen in figure \ref{fig_effmass}. Here, the effective mass has been extracted from DFT data using a fit with the formula given in equation \ref{equ_bands}. Up to 6\% strain, both models agree surprisingly well --- above 6\%, the models differ significantly. At the critical strain, one should consider the contributions from more than one band. Due to the exponential decay of the Fermi function defining the bands' occupation, this only plays a role narrowly around the critical strain. As we do not explicitly focus on this region, this effect can be neglected safely. 

\begin{figure}[t] 
\includegraphics[width=84mm]{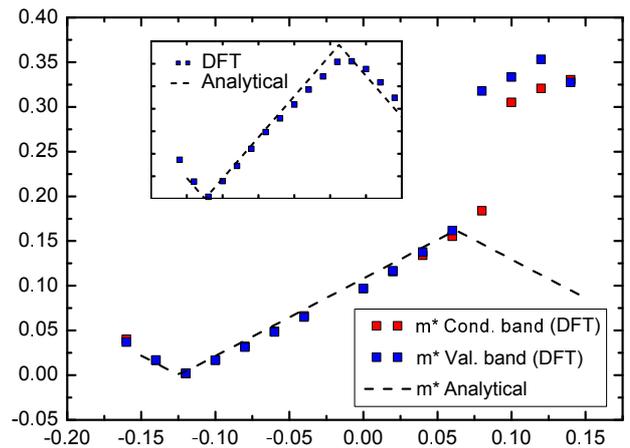}
\caption{Effective mass of carriers: The analytical and the DFT effective masses agree nicely in a certain strain range. Above this range, the contributing CNT bands swap (see fig. \ref{fig_bands_swap}) and thus, the analytical model fails. The inset shows the bandgap with respect to strain (already published in \cite{Wagner_2012b}), which agrees satisfyingly over the whole strain range. } \label{fig_effmass}
\end{figure}

For CNTs with a very small chiral angle, this critical strain is reached at lower strain values --- the lowest value for $\varepsilon^\ast$ is about 4\%. As the sensor regime is mostly within zero to a few percent, the analytical model can be used for CNT sensors, generally.

\begin{figure}[t]
\includegraphics[width=84mm]{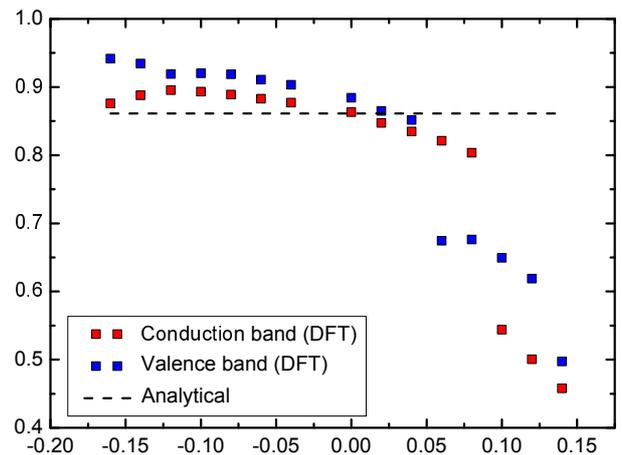}
\caption{Fermi velocity of carriers: Similar to the behavior of the effective mass, the analytical Fermi velocity agrees satisfyingly with the DFT data in the strain range up to 6\%. Above, a strain dependence has to be taken into account.} \label{fig_vFermi}
\end{figure}

The last parameter, the Fermi velocity $v_0$, is not explicitly strain dependent in the analytical model. Figure \ref{fig_vFermi} confirms this for the (8,4)-CNT in the non-critical strain range, where the analytical model is compared to DFT based data. Since $v_0$ is altering for different CNT sub-bands, the Fermi velocity drops instantly when the bands swap at the critical strain $\varepsilon^\ast$ (see figure \ref{fig_bands_swap}). Further, it should be noted that the data in figure \ref{fig_vFermi} show a slight strain dependence of the Fermi velocity also below the critical strain $\varepsilon^\ast$, but it has a minor influence on the transport properties.

In summary, the electronic structure of large CNTs is described correctly for experimentally relevant regimes in the range of zero to at least four percent strain, depending on the CNT's chiral angle. Therefore, we will just make use of the analytical data to discuss the results of the transport model in the following section. 

When the sensor region is leaving the valid region of the analytical model, the dependencies of the parameters can be computed once in advance of the transistor modeling for each CNT-type and they will enter the final model as pre-computed parameter sets. Thus, the transistor model ist still numerically cheap while incorporating the full accuracy of \emph{ab-initio} data.


\section{Electronic transport through strained CNTs}

\begin{figure}[t] 
\centering
\includegraphics[width=84mm]{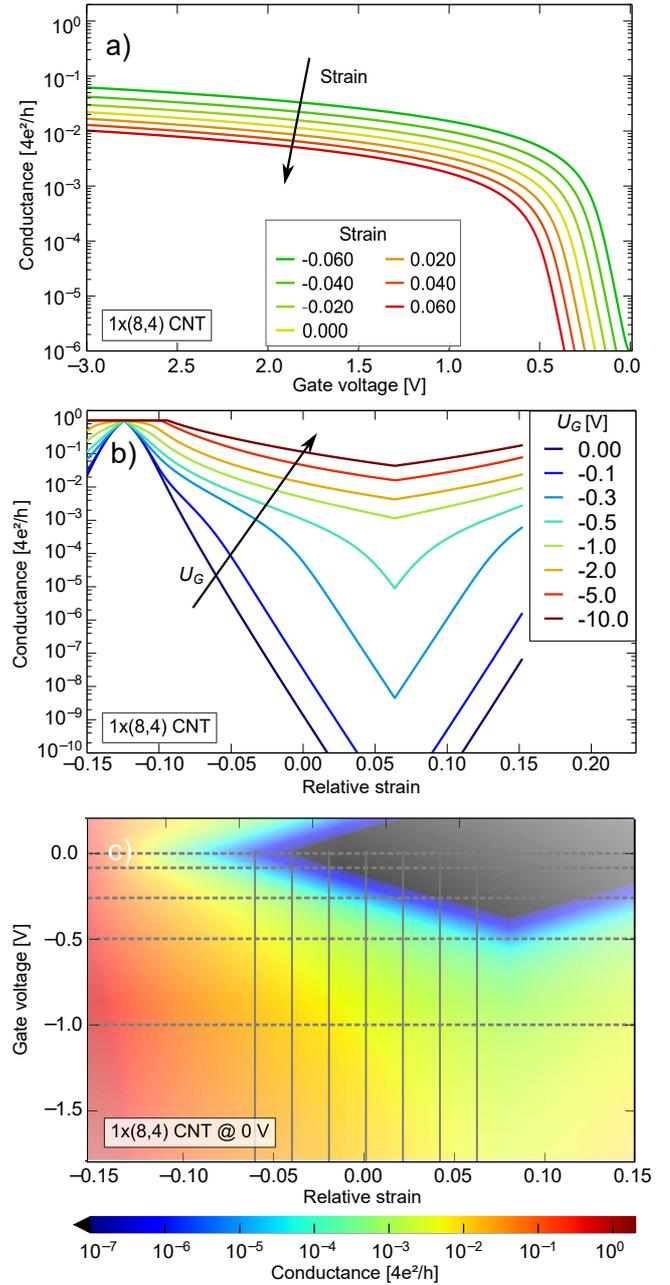} 
\caption{Device characteristics of an (8,4)-CNT within the analytical model -- at varying gate voltages and under strain: We show the transfer characteristics at different strain in a), the strain sensing behavior at different gate voltages in b) and a combined view in c). Horizontal and vertical gray lines in c) are the cuts depicted in a) \& b). The strain region is extended beyond the scope of the model (0--6\% strain) for a more general view.} \label{fig_8_4_CNT}
\end{figure}

The analytical approach described above can now be applied to the calculation of the electronic transport for the (8,4)-CNT in a transistor arrangement. The result is given in figure \ref{fig_8_4_CNT}. 

Subfigure \ref{fig_8_4_CNT}a displays the transfer curves for different strain values. They are similar to theoretical results in \cite{Yoon_2007} and experimental ones in \cite{Muoth_2013} (for comparison see supporting information). The comparison shows that for quantitative predictions, contact effects are important to meet experimental conditions. But the observed trends are very similar. Reference \cite{Ning_2014} shows some interesting results of symmetric transfer characteristics of strained CNTs --- presumably without Schottky-barrier. But somehow, the depicted transfer curves on strain show a $p=1$ behaviour, that does not meet the indicated chirality ((20,12)-CNT, $p=-1$). Knowing that, we cannot compare our model to the data therein.

In our depicted case, the on-current for the (8,4)-CNT decreases strongly with increasing strain which is due to the increasing effective mass $m^\ast$ (see figure \ref{fig_effmass}). In a fully ballistic model without scattering, the strain effect would almost vanish. So, the strain dependence of the device in the on-state (figure \ref{fig_8_4_CNT}a) is dominated by the strain dependence of the effective mass.

Figure \ref{fig_8_4_CNT}b shows the intrinsic CNT conductance as a function of the axial strain for different finite gate voltages. Without gate, the current would be too low to be measured in a real device \footnote{The off-current is underestimated as band-to-band-tunneling is not included. The argumentation is not affected by this.}. Thus, it is needed to raise the overall conductance. On the other hand, the slope of the conductance with respect to strain is decreasing upon increasing gate voltage and the device sensitivity sinks. So, the operation of a real device would be a trade-off between sensitivity and current.

The CNT conductance at zero-bias with respect to strain $\varepsilon$ and applied gate voltage $U\sub{G}$ is depicted in subfigure \ref{fig_8_4_CNT}c. This view shows the strong dependence of the conductance on the gate voltage and the applied strain. The lower the gate voltage, the more sensitive is the CNT device --- but also the conductance is too low to be measured. 

\begin{figure}[t] 
\includegraphics[width=84mm]{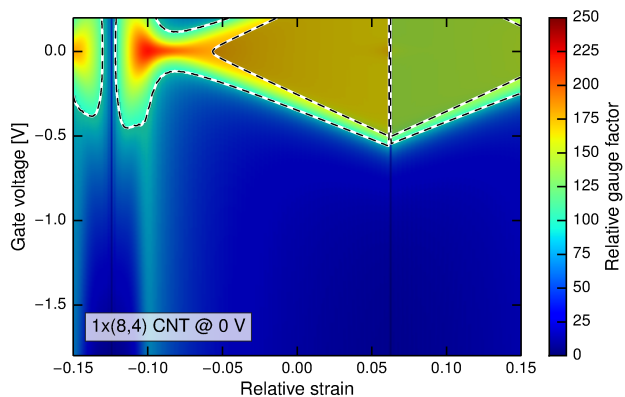} 
\caption{The relative gauge factor $GF$ as a function of strain and gate voltage using the analytical model. The highlighted area shows the preferred working range ($GF > 100$, $G > 10^{-5}\frac{e^2}{h}$) of a CNT strain sensing device based on a (8,4)-CNT. The strain region is extended beyond the scope of the model (0--6\% strain) for a more general view.} \label{fig_8_4_gauge}
\end{figure}

To investigate the sensing region in more detail, figure \ref{fig_8_4_gauge} shows the relative gauge factor in the same parameter region. Mostly, the absolute gauge factor is used as a measure for the strain sensitivity, but this measure is designed for materials whose conductance depends linear on axial strain. The conductance of CNTs depends exponentially on strain such that the relative gauge factor

\begin{equation}
GF = \frac{1}{R}\frac{\partial R(\varepsilon)}{\partial \varepsilon}
\end{equation}

is a better measure for the strain sensitivity. If one defines a device being optimally working at a single CNT conductance\footnote{Depending on the experiment, one uses approximately 100--1000 CNTs instead of a single one.} larger than $10^{-5}e^2/h \approx 4\cdot 10^{-10}\,$S \mbox{(2,5\,$\mathrm{G\Omega}$)} and a relative gauge factor higher than 100, one can find the device operating region for this particular CNT. This region is highlighted in the figure \ref{fig_8_4_gauge}. 

This result is especially helpful, when CNT mixtures are considered for calculation --- which is mostly the reality in experiments. Some key influences are given by metallic and semimetallic CNTs: The admixture of metallic CNTs, for example, strongly reduces the sensitivity in the regime of low gate voltages. The admixture of a small fraction of semimetallic CNTs is affecting the sensor behavior for small strain. A few, more detailed results concerning CNT mixtures can be found in the supporting material.

In total, we described the CNT transistors' strain sensing behavior at room temperature. In a CNT strain sensing device, one has to trade off between sensitivity and magnitude of the signal. Due to the low computational effort required for the evaluation of the analytical expressions, the model can be used for compact modeling of CNT-transistor based strain sensors as well. Furthermore, by including tables of pre-computed results from \emph{ab-inito} data we may obtain compact models covering a very wide range of sensor conditions and CNT types. 

Our model shows some important differences to the model of Minot \cite{Minot_2003}, which is often used for performance estimation of CNT FET strain sensors. The predicted current in their model is proportional to $I_\tn{off}\sim \ee^{\frac{E_G}{k_B T}}$, which would be the maximum influence of the contacts --- the 'transport gap' is always twice the (strained) CNT bandgap. This is reasonable for the experimental situation with a strongly local gate (AFM tip) in the off-state. Basically, this is an approximation for the off-state. In contrast, our model yields $I_\tn{off}\sim \ee^{\frac{E_G}{2 k_B T}}$, as contact effects are not included and the gate is assummed to be homogenous. Our extension contains the dependence on the gate voltage, such that the on-state can be described, too. 

Thus, the suitable model depends on the experimental situation. Usually, the strain dependence of the off-current is only relevant, when the CNT bandgap is small. Band-to-band-tunneling (BTBT)-currents can be larger\cite{Maneux_2013,Luo_2013,Claus_2012}. 

The realistic behaviour demands a rigorous treatment of contact effects, based on e.g. \cite{Leonard_2006} or \cite{Knoch_2008}. Still, in experiments, the exact surface is unknown and even the same contact material (Pd) shows a different CNT transistor behaviour \cite{Ning_2014,Chen_2005,Svensson_2009,Muoth_2013}. In current literature, the contact physics of CNT transistors is not fully clear. 

In our case, we can model BTBT and contact tunneling probabilities in a simplified way --- by introducing by a parallel resistance and a contact tunneling probability. These quantities are fitted to experiments at zero strain, such that more reasonable predictions for a specific sensor are possible (see Supplemental information). 

And even if our model does not contain the explicit influence of strained contacts, the general trends concerning device operating region and strain sensitivity are comparable to experiments in the low bias regime.

\section{Conclusion and perspectives}

The presented results show that the electronic structure of strained CNTs with a radius larger than 0.7\,nm can be described by an analytical model up to critical strain values with almost \emph{ab-initio}-accuracy. The parameters characterizing the electronic structure, especially the critical strain, depend on the CNT's chiral angle. If the quality of the analytical data is not sufficient, parameters from electronic structure calculations can be incorporated. For most sensor applications, the analytical model should be sufficient.

A transport model, based on near-ballistic transport including semiempiric phonon scattering \cite{Zhou_2005}, is described and its results are discussed. Thermal effects and strain are included consistently, which is new in literature. Due to a little amount of available, experimental data of CNTs in a transistor geometry \cite{Minot_2003,Muoth_2013}, the quality of the presented model cannot easily be judged. Although effects of the contacts are not considered in our model so far (e.g. band-to-band-tunneling, non-equilibrium charge carrier distributions), the model predicts trends concerning strain sensitivity and altering of the gate voltage reliably.

Our model predicts that the best sensing behavior can be found for low gate voltages between 0.2 and 0.5 V (with respect to the transistors off-state) and little strain for the (8,4)-CNT, which is desirable for experimental conditions. Still, this range depends on the actual CNT type, admixtures of other CNT chiralities as well as on the demands of the sensor. 

The authors aim the incorporation of the actual model into compact models including contact effects, e.g. \cite{Maneux_2013,Schroter_2013,Luo_2013}. Here, more work on the calculation of \textit{ab-initio} transport models has to be done. The incorporation of contact effects with respect to strain can have a strong influence on the magnitude of the sensor signal of CNTs, so that a thorough \emph{ab-initio} analysis of strained contacts is necessary.

\begin{acknowledgements}

The authors gratefully acknowledge the funding by the German Research Foundation (DFG) through the DFG research unit 1713. We also acknowledge the scientific support by the group of professor Michael Schreiber.

\end{acknowledgements}

\vfill


\end{document}